\documentclass[12pt]{article}
\usepackage{graphicx}
\usepackage{mathtools,slashed}

\newcommand\pubnumber{}
\newcommand\pubdate{\today}

\def\institute{High Energy Physics Institute of\\
Tbilisi State University, Tbilisi, Georgia}
\def\support{\footnote{The author was funded by the grant G-2098 through ISTC and by grants DI/20/6-200/14, 31/44 through SRNSF.}}

\def\Title#1{\begin{center} {\Large #1 } \end{center}}
\def\Author#1{\begin{center}{ \sc #1} \end{center}}
\def\Address#1{\begin{center}{ \it #1} \end{center}}

\newcommand\pubblock{\rightline{\begin{tabular}{l} \pubnumber\\
         \pubdate  \end{tabular}}}
\newenvironment{Abstract}{\begin{quotation}  }{\end{quotation}}
\newenvironment{Presented}{\begin{quotation} \begin{center} 
             PRESENTED AT\end{center}\bigskip 
      \begin{center}\begin{large}}{\end{large}\end{center} \end{quotation}}

\def\beq{\begin{equation}}
\def\eeq#1{\label{#1}\end{equation}}
\def\eeqn{\end{equation}}

\def\beqa{\begin{eqnarray}}
\def\eeqa#1{\label{#1}\end{eqnarray}}
\def\eeqan{\end{eqnarray}}

\let\bar=\overbar

\def\Dslash{\not{\hbox{\kern-4pt $D$}}}
\def\dslash{\not{\hbox{\kern-2pt $\del$}}}

\def\msb{{\bar{\ssstyle M \kern -1pt S}}}

\begin{document}
\begin{titlepage}
\pubblock

\vfill
\Title{Search for $tZ$ Flavour Changing Neutral Currents in top-quark decays with the ATLAS detector}
\vfill
\Author{ A. Durglishvili\support}
\Author{On behalf of the ATLAS Collaboration}
\Address{\institute}
\vfill
\begin{Abstract}
A search for flavour-changing neutral current (FCNC) processes in top-quark decays is presented. Data collected from proton-proton collisions at the LHC at a centre-of-mass energy of $\sqrt{s}=13$ TeV corresponding to an integrated luminosity of 36 fb$^{-1}$, are analysed. A search is performed for top-quark pair-production events, with one top-quark decaying through the $t\to qZ$ ($q=u,c$) FCNC channel, and the other through the dominant Standard Model mode $t\to bW$. Only the decays of the $Z$ boson to charged leptons and leptonic $W$ boson decays are considered as signal. No evidence for a signal is found. The observed and expected upper limits on the branching ratio of $t\to uZ$ and $t\to cZ$ are set at 95\% confidence level and are about a factor 3 better than the ones obtained with the Run 1 data of the ATLAS detector.
\end{Abstract}
\vfill
\begin{Presented}
$10^{th}$ International Workshop on Top Quark Physics\\
Braga, Portugal,  September 17--22, 2017
\end{Presented}
\vfill
\end{titlepage}
\def\thefootnote{\fnsymbol{footnote}}
\setcounter{footnote}{0}

\section{Introduction}
The top quark is the heaviest elementary particle known, with a mass 
$m_t=173.1\pm0.6$ GeV~\cite{Patrignani:2016xqp}.
In the the Standard Model of particle physics (SM), it decays almost exclusively to $bW$ and 
flavour-changing neutral current~(FCNC) decays such as $t\to qZ$ are 
forbidden at tree level. However, FCNC decays occur at one-loop level, but are strongly
suppressed by the GIM mechanism~\cite{Glashow:1970gm} with a suppression factor of 14
orders of magnitude with respect to the dominant decay 
mode~\cite{AguilarSaavedra:2004wm}.  This essentially guarantees
that  any  measurable  branching  ratio (BR) for  top  FCNC  decays  is  an  indication  of  new  physics.  
Several SM extensions predict higher branching ratios for top-quark FCNC decays, see Ref.~\cite{Agashe:2013hma} for a comprehensive review of the various extensions of the SM that have been proposed. 
 
Experimental limits on the FCNC $t\to qZ$ BR were established by several experiments~\cite{Heister:2002xv,Abdallah:2003wf, Abbiendi:2001wk,
Achard:2002vv,LEP-Exotica-WG-2001-01,Abramowicz:2011tv,
Aaltonen:2008ac,Abazov:2011qf,CMS-TOP-12-039,TOPQ-2014-08}.
Before the results reported here the most stringent limits, 
BR($t\to uZ$) $<2.2\times 10^{-4}$ and BR($t\to cZ$) $<4.9\times 10^{-4}$ at 95\% 
confidence level, were the ones 
from the CMS Collaboration~\cite{CMS-TOP-12-039} using data collected at 
$\sqrt{s}=8$ TeV. For the same centre-of-mass energy, the ATLAS 
Collaboration derived a limit of BR($t\to qZ$) $<7\times 10^{-4}$~\cite{TOPQ-2014-08}.

This analysis presents a search for the FCNC decay
$t\to qZ$ in top-quark--top-antiquark events with one top quark decaying through the
FCNC mode and the other through the dominant SM mode ($t\to bW$). Only the 
decays of the $Z$ boson into charged leptons\footnote{In this article, lepton denotes electron or muon, including those coming from leptonic tau decays.} and leptonic $W$ boson decays 
are considered. The final-state topology is thus characterised by the 
presence of three isolated charged leptons, at least two jets with exactly one being tagged as a jet containing $b$-hadrons, and missing transverse momentum from the undetected neutrino. 

\section{Samples and selection}
This analysis uses $pp$ collisions data collected by the
ATLAS detector during 2015 and 2016 at a centre-of-mass
energy of $\sqrt{s}=13$ TeV and corresponding to an
integrated luminosity of $L = 36$ fb$^{-1}$.
Monte-Carlo simulation samples (MC) are used to model the signal and several background processes.
The main source of background events containing three 
real leptons are diboson production, $t\bar{t}Z$ and $tZ$ processes. These backgrounds are estimated using the MC simulation. An additional 
background originates from events with two or fewer real leptons and 
additional non-prompt\footnote{Prompt leptons are leptons from the decay of real $W$ or $Z$ bosons, either
directly or through the chain of $\tau\to \ell\nu\nu$, or from the
semileptonic decay of top-quarks.} leptons. Such a background is estimated using the MC samples corrected by data in the dedicated control regions.

According to the signal topology events are requied to contain exactly three 
isolated charged leptons with $|\eta|<2.5$ and $p_{\text{T}} > 15$ GeV. The $Z$ boson candidate is reconstructed from the two leptons that 
have the same flavour, opposite charge and a reconstructed mass within 
15 GeV of the $Z$ boson mass ($m_Z$~\cite{Patrignani:2016xqp}).
If more than one compatible lepton-pair is found, the one with the reconstructed mass closest 
to $m_Z$ is chosen as the $Z$-boson candidate. According to the signal 
topology, the events are then required to have $E_{\text{T}}^{\text{miss}}>20$ GeV and at least 
two jets. All jets are required to have $p_{\text{T}} > 25$ GeV and $|\eta|<2.5$. 
Exactly one of the jets must be $b$-tagged.
Applying energy--momentum conservation, the kinematic properties of the top quarks 
are reconstructed from the corresponding decay particles by minimising, without constraints, the 
following expression:
\begin{equation}
\chi^2  =  \frac{\left(m^{\mathrm{reco}}_{j_a\ell_a\ell_b}-m_{t_{\mathrm{FCNC}}}\right)^2}{\sigma_{t_{\mathrm{FCNC}}}^2}+
\frac{\left(m^{\mathrm{reco}}_{j_b\ell_c\nu}-m_{t_{\mathrm{SM}}}\right)^2}{\sigma_{t_{\mathrm{SM}}}^2}
+ \frac{\left(m^{\mathrm{reco}}_{\ell_c\nu}-m_W\right)^2}{\sigma_W^2},
\label{eq:chi2}
\end{equation}
where $m^{\mathrm{reco}}_{j_a\ell_a\ell_b}$, 
$m^{\mathrm{reco}}_{j_b\ell_c\nu}$ and $m^{\mathrm{reco}}_{\ell_c\nu}$ are 
the reconstructed masses of the $qZ$, $bW$ and $\ell\nu$ systems, 
respectively. 
For each jet combination $j_b$ 
must correspond to the $b$-tagged jet, while any jet can be assigned to $j_a$. 
The $E_{\text{T}}^{\text{miss}}$ is assumed to be the transverse momentum of the neutrino in the $W$ boson decay.
The longitudinal component of the neutrino momentum ($p^{\nu}_z$) is then determined
by the minimisation of Eq. \ref{eq:chi2}. The $\chi^2$ minimisation gives the most probable value for $p^{\nu}_z$. From all combinations, the one with the  minimum $\chi^2$ is chosen, along with the corresponding $p^{\nu}_z$ value. 

The final requirements to define the signal region (SR) are 
$|m^{\mathrm{reco}}_{j_a\ell_a\ell_b} - 172.5~\text{GeV}|<40$ GeV, 
$|m^{\mathrm{reco}}_{j_b\ell_c\nu} - 172.5~\text{GeV}|<40$ GeV and 
$|m^{\mathrm{reco}}_{\ell_c\nu} - 80.4~\text{GeV}|<30$ GeV. 
\section{Statistical analysis}
The statistical analysis to extract the signal is based on a binned likelihood function 
$L(\mu,\theta)$ constructed as a product of Poisson probability terms over 
all bins in each considered distribution, and several Gaussian constraint 
terms for $\theta$, a set of nuisance parameters that parametrise effects of 
statistical uncertainty and all sources of systematical uncertainties (theoretical, experimental and event modelling) on the signal and 
background expectations. This function depends on the signal-strength 
parameter $\mu$, a multiplicative factor for the number of signal events 
normalised to a reference branching ratio. The distributions from the SR and pre-defined
background control regions (for non-prompt leptons, $WZ$, $ZZ$ and $t\bar{t}Z$ events) are combined to test for the presence of a signal. The $\chi^2$ variable from the reconstruction of top-quark candidates is used for the signal-to-background discrimination in the SR. No attempt is made to separate the signal from the
background in the background control regions, but they allow a tighter constraint of backgrounds and systematic uncertainties in a combined fit with the signal region. 

The test statistic $q_{\mu}$ is defined as the profile likelihood ratio~\cite{Cowan:2010js}: 
\begin{equation*}
q_{\mu}=-2\ln(L(\mu,\hat{\hat{\theta}}_{\mu})/L(\hat{\mu},\hat{\theta})),
\end{equation*}
which is used to measure the compatibility of the observed data with the background-only hypothesis
(i.e. for $\mu = 0$), and to make statistical inferences about $\mu$, such as upper limits using the CL$_s$ method~\cite{Read:2002hq, Junk:1999kv}.
\section{Results}
\begin{figure}[htbp]
	\centering
	\includegraphics[width=.4\textwidth]{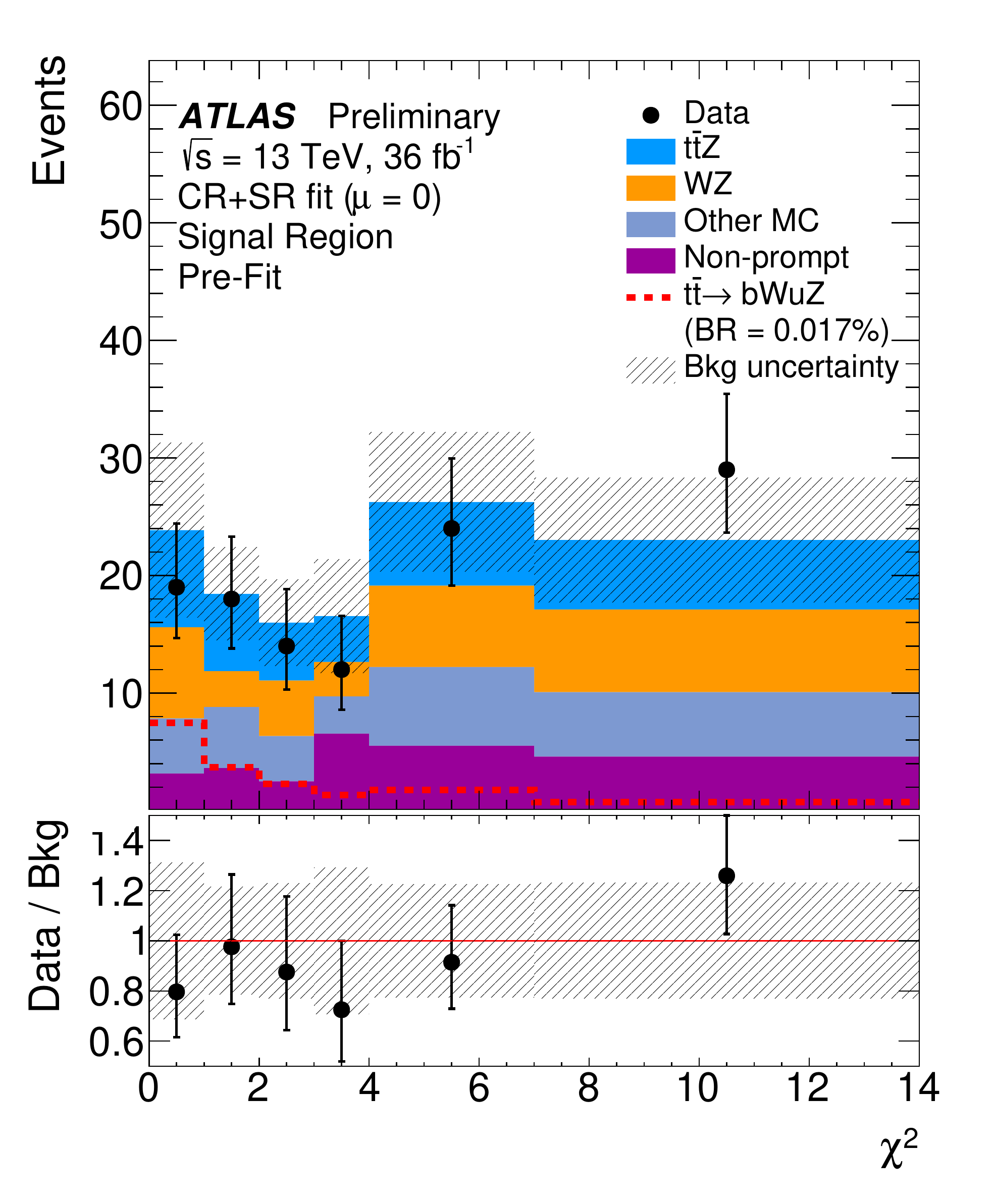} 
	\includegraphics[width=.4\textwidth]{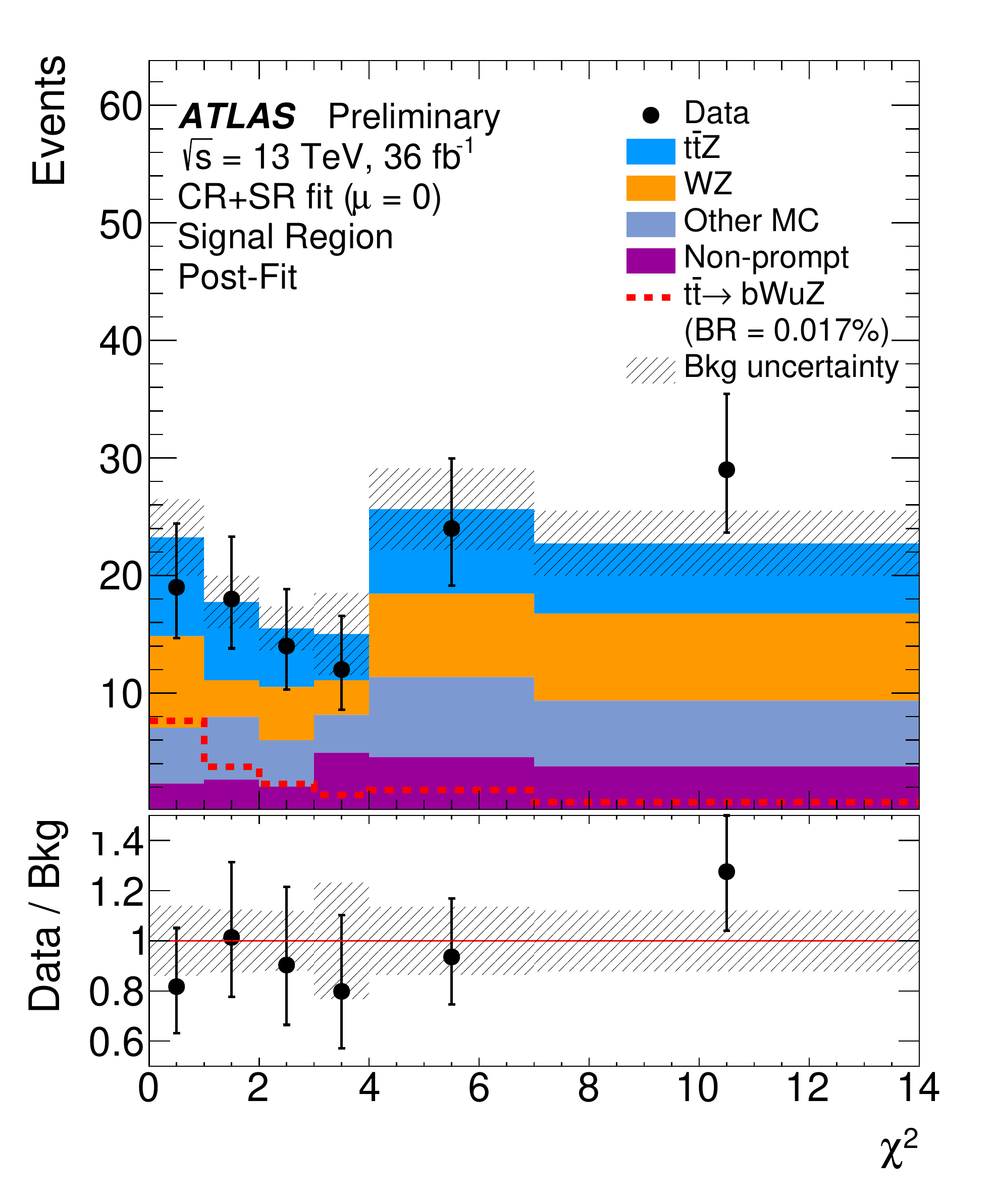} 
	\caption{Expected (filled histogram) and observed (points with error bars) distributions of the $\chi^2$ of the kinematical reconstruction in the SR, before (left) and after (right) the combined fit under the background-only hypothesis.
	For comparison, distributions for the FCNC $t\bar{t}\to bWuZ$ signal (dashed line), normalised to
	the observed limit, are also shown. The dashed area represents the total uncertainty on the background prediction.}
	\label{fig:chi2}
\end{figure}

Good agreement between data and expectation from the background-only hypothesis
is observed and no evidence for an FCNC signal is found. The distribution of the discriminant variable in the SR is presented in Figure~\ref{fig:chi2}, before and after the fit under the background-only hypothesis.
The upper limits on BR$(t\to qZ)$ are computed with the CL$_{s}$ 
method using the asymptotic properties of 
$q_{\mu}$~\cite{Cowan:2010js} and assuming that only one FCNC mode contributes. The resulting observed (expected) limits at 95\% CL are~\cite{ATLAS-CONF-2017-070}:
BR($t\to uZ$) $<1.7\times10^{-4}$ ($2.4\times10^{-4}$) and BR($t\to cZ$) $<2.3\times10^{-4}$ ($3.2\times10^{-4}$).


\section{Conclusions}
A search for FCNC processes in top-quark decays is
presented. The ATLAS data of 13 TeV corresponding to
an integrated luminosity of 36 fb$^{-1}$, are analysed. No
evidence for a signal is found and the observed
(expected) upper limits on the $t\to uZ$ and $t\to cZ$ branching
ratios of $1.7\times10^{-4}$ ($2.4\times10^{-4}$) and $2.3\times10^{-4}$ ($3.2\times10^{-4}$) are
set at 95\% confidence level, respectively.


\begin{thebibliography}{99}


\bibitem{Patrignani:2016xqp}
C. Patrignani et al., Chin. Phys. {\bf C40} (2016) 100001.

\bibitem{Glashow:1970gm}
S. Glashow, J. Iliopoulos and L. Maiani, Phys. Rev. D {\bf 2} (1970) 1285.

\bibitem{AguilarSaavedra:2004wm}
J. Aguilar-Saavedra, Acta Phys. Polon. B {\bf 35} (2004) 2695.

\bibitem{Agashe:2013hma}
Snowmass Top Quark WG, K. Agashe et al., arXiv: 1311.2028 [hep-ph].

\bibitem{Heister:2002xv}
ALEPH Collaboration, A. Heister et al., Phys. Lett. B {\bf 543} (2002) 173.

\bibitem{Abdallah:2003wf}
DELPHI Collaboration, J. Abdallah et al., Phys. Lett. B {\bf 590} (2004) 21.

\bibitem{Abbiendi:2001wk}
OPAL Collaboration, G. Abbiendi et al., Phys. Lett. B {\bf 521} (2001) 181.

\bibitem{Achard:2002vv}
L3 Collaboration, P. Achard et al., Phys. Lett. B {\bf 549} (2002) 290.

\bibitem{LEP-Exotica-WG-2001-01}
The LEP Exotica WG, LEP-Exotica-WG-2001-01, 2001, url: https://cds.cern.ch/record/1006392.

\bibitem{Abramowicz:2011tv}
ZEUS Collaboration, H. Abramowicz et al., Phys. Lett. B {\bf 708} (2012) 27.

\bibitem{Aaltonen:2008ac}
CDF Collaboration, T. Aaltonen et al., Phys. Rev. Lett. {\bf 101} (2008) 192002.

\bibitem{Abazov:2011qf}
D0 Collaboration, V. M. Abazov et al., Phys. Lett. B {\bf 701} (2011) 313.

\bibitem{CMS-TOP-12-039}
CMS Collaboration, JHEP {\bf 07} (2017) 003.

\bibitem{TOPQ-2014-08}
ATLAS Collaboration, Eur. Phys. J. C {\bf 76} (2016) 12.

\bibitem{Cowan:2010js}
G. Cowan, K. Cranmer, E. Gross and O. Vitells, Eur. Phys. J. {\bf C71} (2011) 1554, Erratum: Eur. Phys. J. {\bf C73} (2013) 2501.

\bibitem{Read:2002hq}
A. L. Read, J. Phys. G {\bf 28} (2002) 2693.

\bibitem{Junk:1999kv}
T. Junk, Nucl. Instrum. Meth. {\bf A 434} (1999) 435.

\bibitem{ATLAS-CONF-2017-070}
ATLAS Collaboration, ATLAS-CONF-2017-070, url: https://cds.cern.ch/record/2285808.

\end{thebibliography}
\end{document}